\documentclass[pra,aps,twocolumn,amsmath,amssymb,superscriptaddress]{revtex4-1}
\usepackage{epsfig,amsmath}
\usepackage{subfigure}
\usepackage{graphicx}
\usepackage{dcolumn}
\usepackage{stmaryrd}
\usepackage{mathrsfs}
\usepackage{pifont}
\usepackage{amsthm}
\usepackage{amssymb}
\usepackage{bm}
\usepackage{latexsym}
\usepackage[colorlinks=true,linkcolor=cyan,citecolor=cyan]{hyperref}
\usepackage{color}
\usepackage{epstopdf}
\usepackage[ruled]{algorithm2e}

\begin{document}

\title{Charging by quantum measurement}

\author{Jia-shun Yan}
\affiliation{School of Physics, Zhejiang University, Hangzhou 310027, Zhejiang, China}

\author{Jun Jing}
\email{Email address: jingjun@zju.edu.cn}
\affiliation{School of Physics, Zhejiang University, Hangzhou 310027, Zhejiang, China}

\date{\today}

\begin{abstract}
We propose a quantum charging scheme fueled by measurements on ancillary qubits serving as disposable chargers. A stream of identical qubits are sequentially coupled to a quantum battery of $N+1$ levels and measured by projective operations after joint unitary evolutions of optimized intervals. If charger qubits are prepared in excited state and measured on ground state, then their excitations (energy) can be near-perfectly transferred to battery by iteratively updating the optimized measurement intervals. Starting from its ground state, the battery could be constantly charged to an even higher energy level. Starting from a thermal state, the battery could also achieve a near-unit ratio of ergotropy and energy through less than $N$ measurements, when a population inversion is realized by measurements. If charger qubits are prepared in ground state and measured on excited state, useful work extracted by measurements alone could transform the battery from a thermal state to a high-ergotropy state before the success probability vanishes. Our operations in charging are more efficient than those without measurements and do not invoke the initial coherence in both battery and chargers. Particularly, our finding features quantum measurement in shaping nonequilibrium systems.
\end{abstract}

\maketitle

\section{Introduction}

Over one century, the classical batteries have been driving the revolutions in personal electronics and automotive sector. As energy-storage units in a cutting-edge paradigm, quantum batteries~\cite{QuantumVSClassical,QuantumFlywheel,CapacityPowerBouonds,EntanglementNotNecessary,CapacityPowerBouonds} are expected to outperform their classical counterparts by widely exploiting the advantages from quantum operations and promoting their efficiency under the constraint of quantum thermodynamics~\cite{EntanglementEnergyExtract,WorkAndCorrelation,FundamentaLlimitation}. Enormous attentions were paid to charging quantum batteries, as the primary step in the charge-store-discharge cycle. Many protocols have been proposed, including but not limited to charging by entangling operations~\cite{EnhancingChargingPower,CannotWithoutGlobalOperation}, charging with dissipative~\cite{DissipativeCharging,ThermalizationCharging}, unitary, and collision processes~\cite{ChargerMediatedEnergyTransfer,CollisionalCharging}, charging collectively and in parallel~\cite{CollectiveCharging,Quantacell}, and charging with feedback control~\cite{CharingbyFeedback}. Many-body interaction~\cite{SpinChainBattery,ManybodyBattery,SYKCharging} and energy fluctuation~\cite{ChargingPowerBoundedByFluctuation,RandomQuantumBattery} were also explored to raise the upper-bound of charging power and capacity. Local and global interactions are designed to transfer energy from various thermodynamical resources to batteries. The maximum rate and amount in energy transfer are subject to relevant timescales of evolution and relaxation.

Quantum measurements, particularly the repeated projections onto a chosen state or a multidimensional subspace, could change dramatically the transition rate of the measured system~\cite{ZenoParadox,QuantumZeno,ZenoSubspace}. Numerous measurement-based control schemes were applied to state purification~\cite{ContinuousMeasurePurify,QubitPurification}, information gain~\cite{MaximumInfGain}, and entropy production~\cite{EntropyProductionBelenchia,EntropyProductionLandi}. Quantum engineering by virtue of the measurements on ancillary system, that generates a net nonunitary propagator, is capable to purify and cool down quantum systems~\cite{SimultaneousCooling,Purification,MeasurementCoolingWu,MeasurementCoolingExp,CoolingEngine}. In general, a projective measurement or postselection on ancillary system would navigate the target system to a desired state with a finite probability. Therefore, quantum measurements could become a useful resource as well as the heat or work reservoirs, serving as fuels powering a thermodynamical or state-engineering scheme through a nonunitary procedure~\cite{MeasurementEnergetics,GateEnergetics,MeasurementThermodynamics,MeasurementInThermodynamics,MeasurementEngines}. In this work, we address quantum measurements in the context of quantum energetics by the positive operator-valued measures (POVM). It is interesting to find that POVMs generated by the joint evolution of chargers and batteries combined with projections on a specific state of chargers is able to speed up the charging rate and promote the amount of accumulated energy and ergotropy. Our method is transparently distinct from those based on swap or exchange operations. It does not necessarily rely on the initial states of both battery and charger. Without energy exchange, it could transform the system from a completely passive state to a useful state for battery.

\begin{figure}[htbp]
\centering
\includegraphics[width=0.95\linewidth]{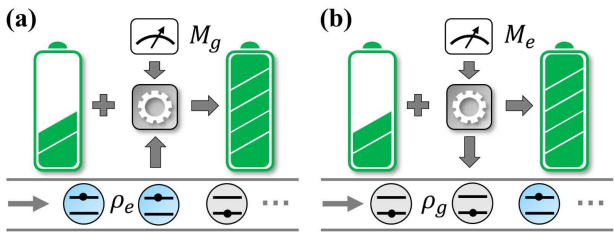}
\caption{(a) Power-on and (b) Power-off charging schemes. Identical ancillary qubits line up to interact with the battery for a period of time. In the end of each round of joint evolution, a projective measurement is performed on the qubit to induce a nonunitary charging operation on the battery. In (a) power-on charging, qubits are initialized as the excited state $\rho_e=|e\rangle\langle e|$ and the projection $M_g=|g\rangle\langle g|$ is performed on its ground state. Energy gain for the battery is mainly from the qubit excitations. In (b) power-off charging, qubits are initialized as the ground state $\rho_g=|g\rangle\langle g|$ and the projection $M_e=|e\rangle\langle e|$ is on the excited state. Useful work extracted by measurements simultaneously charges both battery and ancillary qubits.}\label{Model}
\end{figure}

In particular, we propose a charging-by-measurement scheme in a quantum collision framework~\cite{QuanMechOfMeasurement,PositionMeasurement,OpenSystemModeling}. As disposable chargers a sequence of identical qubits line up to temporarily interact with the battery (a multilevel system with a finite number of evenly spaced ladders). Once a projective measurement is performed in the end of the joint evolution and the outcome is as desired, the coupled qubit is replaced with a new one and then the charging continues. Figures~\ref{Model}(a) and \ref{Model}(b) demonstrate respectively a {\em power-on} and a {\em power-off} charging schemes. When the qubits are initially in the excited state $\rho_e=|e\rangle\langle e|$, the projective measurement on the ground state $M_g=|g\rangle\langle g|$ transfers the energy of excited qubits to the battery. The energy gain for battery increases linearly with the number of measurements and the charging power is gradually enhanced as well. Full population inversion is realized when the measurement number is close to the battery size. When the qubits are prepared as the ground states $\rho_g=|g\rangle\langle g|$, still the battery can be charged by measuring the ancillary qubits on the excited state $M_e=|e\rangle\langle e|$. Useful work extracted entirely from repeated measurements simultaneously charges both battery and charger.

The rest of this work is structured as follows. In Sec.~\ref{MeasureFueledCharger}, we introduce a general model of charging by measurements, whereby the charging or discharging effect is analyzed in view of a general POVM. In Sec.~\ref{PwonChargingSec}, we present the power-on charging scheme. We find an optimized measurement interval to maximize the measurement probability, by which both energy and ergotropy of the battery scale linearly with the number of measurements. Section~\ref{PwoffChargingSec} devotes to the power-off charging scheme. For both schemes, we evaluated the charging efficiency by charging power, state distribution, energy and ergotropy of the battery. In Sec.~\ref{Discussion}, we discuss the effect from the initial coherence in the chargers and the robustness of our scheme against the environmental decoherence. In Sec.~\ref{Conclusion}, we summarize the whole work.

\section{General Model of charging by measurements}\label{MeasureFueledCharger}

We aim for charging a quantum battery by performing measurements on the ancillary qubits as disposable chargers. The scheme is constructed by rounds of joint evolution and projective measurements. The full Hamiltonian $H=H_B+H_C+V$ consists of a target battery system $H_B$, a sequence of identical charger qubits (in each round only a single charger with a free Hamiltonian $H_C$ is coupled to the battery and the others are decoupled), and the interaction $V$ between battery and the current working charger. The battery is assumed to be in a thermal state $\rho_B=e^{-\beta H_B}/{\rm Tr}[e^{-\beta H_B}]$ with $\beta\equiv1/k_BT$. It is a completely passive state that is energetic but has no ergotropy. In quantum battery, ergotropy is defined as $\mathcal{E}\equiv{\rm Tr}[H_B(\rho-\sigma)]$, where $\sigma$ is the passive state obtained by realigning the eigenvalues of $\rho$ in decreasing order, which has none extractable energy under cyclic unitary operations~\cite{Ergotropy}. For $\rho=\rho_B$, it is found that $\sigma=\rho$. The thermal state is thus a reasonable choice to demonstrate the power of any charging scheme, also it is a natural state for the battery subject to a thermal bath in the absence of active controls. The charger qubits are prepared as the same mixed state $\rho_C=q|g\rangle\langle g|+(1-q)|e\rangle\langle e|$ with a ground-state occupation $q\in[0, 1]$ before linking to the battery. The initial coherence of chargers is temporally omitted to distinguish the charging efficiency of quantum measurements.

In each round, the joint evolution of the battery and the working qubit is described by the time-evolution operator $U=\exp(-iH\tau)$. The coupling interval $\tau$ might be constant or vary with respect to all the rounds. An instantaneous projective measurement $M$ on the qubit is implemented in the end of the round, and then the (unnormalized) joint state becomes
\begin{equation}\label{joint}
\rho_{\rm tot}'=MU\rho_B\otimes\rho_CU^\dagger M
\end{equation}
with a finite measurement probability $P={\rm Tr}[MU\rho_B\otimes\rho_CU^\dagger]$. In this work, we do not consider the errors occurring in measurements and its energy cost. After measurement, the charger qubit is decoupled from the battery system and withdrawn, then another one is loaded to the next round. The charging scheme is nondeterministic in essence and thus employs a feedback mechanism: the measurement outcome determines whether to launch the next round of charging cycle or to restart from the beginning.

The quantum battery in our model has $N+1$ energy ladders with Hamiltonian $H_B=\omega_b\sum_{n=0}^Nn|n\rangle\langle n|$, where $\omega_b$ is the energy unit of the battery. $\hbar=k_B=1$. The ladder operators are defined as $A^\dagger=\sum_{n=1}^N\sqrt{n}|n\rangle\langle n-1|$ and $A=\sum_{n=1}^N\sqrt{n}|n-1\rangle\langle n|$. The Hamiltonian for each charge qubit is $H_C=\omega_c|e\rangle\langle e|$, where $\omega_c$ is the energy spacing between the ground $|g\rangle$ and excited states $|e\rangle$. Battery and qubits are coupled with the exchange interaction $V=g(\sigma_-A^\dagger+\sigma_+A)$, where $g$ is the coupling strength and $\sigma_-$ and $\sigma_+$ denote the transition operators of the qubit. Then the full Hamiltonian in the rotating frame with respect to $H_0=\omega_b(\sum_{n=0}^Nn|n\rangle\langle n|+|e\rangle\langle e|)$ reads
\begin{equation}\label{Ham}
H=\Delta|e\rangle\langle e|+g\left(\sigma_-A+\sigma_+A^\dagger\right),
\end{equation}
where $\Delta=\omega_c-\omega_b$ represents the energy detuning between charger qubit and battery.

The charging procedure is piecewisely concatenated by a sequence of joint evolutions of charger qubit and battery, which is interrupted by instantaneous projective measurements $M_\varphi=|\varphi\rangle\langle\varphi|$ over a particular state $|\varphi\rangle=\cos(\theta/2)|g\rangle+\sin(\theta/2)|e\rangle$, $0\leq\theta\leq\pi$, of the working qubit. After a round with an interval $\tau$, the battery state becomes
\begin{equation}\label{rhoB}
\rho_B(\tau)={\rm Tr}_C\left[\rho_{\rm tot}'\right]=\frac{\mathcal{D}+\mathcal{C}}{P_{\varphi}}=\frac{\mathcal{D}+\mathcal{C}}{{\rm Tr}[M_{\varphi}U\rho_B\otimes\rho_CU^\dagger]},
\end{equation}
where $\mathcal{D}$ is the diagonal (population) part in the density matrix of the battery system without normalization and $\mathcal{C}$ is the off-diagonal part
\begin{equation}\label{DynCoherence}
\mathcal{C}=\frac{\sin\theta}{2}\sum_{n=1}^N\alpha_n^*\left[q\lambda_np_n+(1-q)\lambda_n^*p_{n-1}\right]|n-1\rangle\langle n|+{\rm H.c.}
\end{equation}
Here both $\alpha_n=\cos\Omega_n\tau+i\Delta\sin(\Omega_n\tau)/2\Omega_n$ and $\lambda_n=-ie^{-i\Delta\tau/2}g\sqrt{n}\sin(\Omega_n\tau)/\Omega_n$ are renormalization coefficients, $\Omega_n=\sqrt{g^2n+\Delta^2/4}$ is the Rabi frequency, and $p_n$ is the initial thermal occupation on the $n$th level of battery, i.e., $\rho_B=\sum_{n=0}^Np_n|n\rangle\langle n|$ with $p_n=[e^{-\beta \omega_bn}-e^{-\beta \omega_b(n+1)}]/[1-e^{-\beta \omega_b(N+1)}]$. $\mathcal{C}$ describes the dynamical coherence that appears during the joint evolution of charger and battery, generating nonzero extractable work for the battery~\cite{CoherenceAndExtractableWork}, and disappears upon the projective measurements.

The population part of the battery state could be divided as
\begin{equation}\label{diag}
\mathcal{D}=\mathcal{D}_{\rm charge}+\mathcal{D}_{\rm discharge}
\end{equation}
due to the heating or cooling contribution on the battery system from various POVMs. By Eqs.~(\ref{joint}) and (\ref{rhoB}), we have
\begin{equation}\label{Population}
\begin{aligned}
&\mathcal{D}_{\rm charge}=(1-q)\cos^2\frac{\theta}{2}\mathcal{M}_{eg}[\rho_B]+q\sin^2\frac{\theta}{2}\mathcal{M}_{ge}[\rho_B], \\
&\mathcal{D}_{\rm discharge}=q\cos^2\frac{\theta}{2}\mathcal{M}_{gg}[\rho_B]+(1-q)\sin^2\frac{\theta}{2}\mathcal{M}_{ee}[\rho_B],
\end{aligned}
\end{equation}
where $\mathcal{M}_{ij}[\cdot]\equiv R_{ij}\cdot R_{ij}^{\dagger}$, $i, j\in\{e, g\}$, represents the $j$th element of the $i$th POVM, satisfying the normalization condition $\sum_jR_{ij}^{\dagger}R_{ij}=I_B$. $R_{ij}\equiv\langle j|U|i\rangle$ is the Kraus operator acting on the state space of the battery, where $i$ and $j$ label respectively the initial state and the measured state of the ancillary qubit. In particular, we have
\begin{equation}\label{POVM}
\begin{aligned}
&\mathcal{M}_{eg}[\rho_B]=\sum_{n=1}^N|\lambda_{n}(\tau)|^2p_{n-1}|n\rangle\langle n|,\\
&\mathcal{M}_{ge}[\rho_B]=\sum_{n=0}^{N-1}|\lambda_{n+1}(\tau)|^2p_{n+1}|n\rangle\langle n|,\\
&\mathcal{M}_{gg}[\rho_B]=\sum_{n=0}^N|\alpha_{n}(\tau)|^2p_n|n\rangle\langle n|,\\
&\mathcal{M}_{ee}[\rho_B]=\sum_{n=0}^{N-1}|\alpha_{n+1}(\tau)|^2p_n|n\rangle\langle n|.
\end{aligned}
\end{equation}

According to Naimark's dilation theorem~\cite{Naimark}, a set of projective measurements $\{M_j\}$ acting on one of the subspaces $\mathcal{H}_1$ of the total space $\mathcal{H}_{\rm tot}=\mathcal{H}_1\otimes\mathcal{H}_2$ could induce a POVM on another subspace $\mathcal{H}_2$ with a map $\mathcal{H}_2\rightarrow\mathcal{H}_{\rm tot}$. In our context, an arbitrary projective measurement $M_{\varphi}=|\varphi\rangle\langle\varphi|$ defined in the space of the charger qubit induces a POVM $\mathcal{M}_{i,\varphi}[\rho_B]$ acting on the battery. And the map is constructed by $U|i\rangle$ with the joint unitary evolution $U$, according to the initial state of the charger $|i\rangle$. Therefore, as shown in Eq.~(\ref{POVM}), we have two sets of POVMs in the bare basis of the charger qubit: $\mathcal{M}_{e,j\in\{e,g\}}$ and $\mathcal{M}_{g,j\in\{e,g\}}$. For instance, $\mathcal{M}_{eg}$ represents a POVM on the battery induced by projection on the ground state of the qubit that is initially in the excited state.

\begin{figure}[htbp]
\centering
\includegraphics[width=0.95\linewidth]{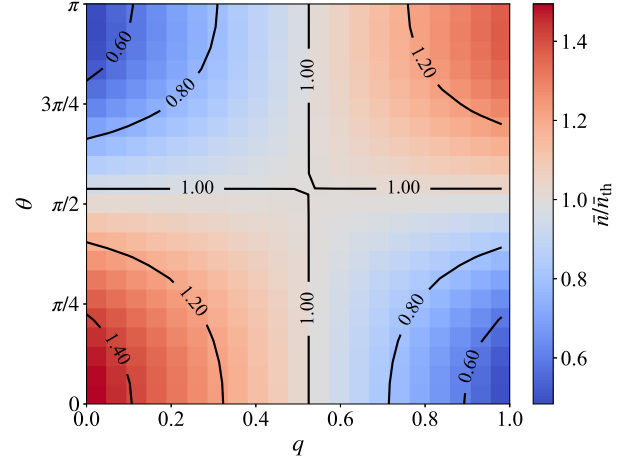}
\caption{Ratio $\bar{n}/\bar{n}_{\rm th}$ of the average population $\bar{n}$ of the battery after a single measurement and the initial thermal population $\bar{n}_{\rm th}\equiv{\rm Tr}[\hat{n}\rho_B]$ in the space of $\theta$ and $q$. The battery size is $N=100$, the initial inverse temperature is $\beta=0.1/\omega_c$, and the measurement interval is $\tau=8/\omega_c$. }\label{ChargingDischarging}
\end{figure}

$\mathcal{D}_{\rm charge}$ in Eq.~(\ref{Population}) is a linear combination of $\mathcal{M}_{eg}$ and $\mathcal{M}_{ge}$ in Eq.~(\ref{POVM}). Under a proper $\tau$, $\mathcal{M}_{eg}$ replaces a smaller $p_n$ with a larger $|\lambda_n|^2p_{n-1}$ for all the excited states. On the contrary, $\mathcal{M}_{ge}$ moves the populations on higher levels to lower levels, which might also enhance the battery energy through a significant renormalization over the population distribution. Note the largest population $p_0$ on the ground state has been eliminated. In contrast, $\mathcal{D}_{\rm discharge}$ is a linear combination of $\mathcal{M}_{gg}$ and $\mathcal{M}_{ee}$. Both of them reduce the populations of the excited state due to the fact that $|\alpha_{n>0}|\leq|\alpha_0|=1$ and then enhance the relative weight of the ground-state population~\cite{MeasurementCoolingWu}. Then they are inclined to discharge the battery.

The dependence of charging or discharging on the initial and measured states can be quantitatively justified in Fig.~\ref{ChargingDischarging} by the ratio of the average population of the charger $\bar{n}$ after a single round of evolution-and-measurement and the initial thermal average population $\bar{n}_{\rm th}$ in the parametric space of $\theta$ and $q$, describing respectively the weights of the measured state and the initial state of qubit on the ground state. The two blue-diagonal corners in Fig.~\ref{ChargingDischarging} correspond to the POVMs $\mathcal{M}_{gg}$ and $\mathcal{M}_{ee}$ that could be used to cool down the target system. For instance, the lower right corner $\mathcal{M}_{gg}$ with $q=1$ and $\theta=0$ describes the mechanism of cooling-by-measurement in the resonator system~\cite{MeasurementCoolingWu}. More crucial to the current work, the lower left corner $\mathcal{M}_{eg}$ with $q=0$ and $\theta=0$ and the upper right corner $\mathcal{M}_{ge}$ with $q=1$ and $\theta=\pi$ motivate our investigation on the following power-on and power-off charging schemes, respectively. One can find that the former scheme is more efficient than the latter in terms of the ratio $\bar{n}/\bar{n}_{\rm th}$ with certain measurement interval.

\section{Power-on charging}\label{PwonChargingSec}

In this section, we present the power-on charging scheme described by $\mathcal{M}_{eg}$ in which the charger qubits are prepared in their excited states with $q=0$ and the projective measurement is performed on the ground state with $\theta=0$. Then the density matrix of the battery after $m\geq1$ rounds of measurements reads
\begin{equation}\label{PwOnChargingrhoB}
\rho_{\rm on}^{(m)}=\frac{\mathcal{M}_{eg}\left[\rho_{\rm on}^{(m-1)}\right]}{P_g(m)}=\frac{\sum_{n=m}^N|\lambda_{n}(\tau)|^2p_{n-1}^{(m-1)}|n\rangle \langle n|}{P_g(m)}
\end{equation}
where $\rho_{\rm on}^{(m-1)}=\sum_{n=m-1}^Np_n^{(m-1)}|n\rangle\langle n|$ denotes the battery state with population $p_n^{(m-1)}$ on the state $|n\rangle$ after $(m-1)$ rounds of measurements under the power-on charging scheme. $p_n^{(0)}=p_n$ describes the initial thermal occupation and $\rho_{\rm on}^{(0)}=\rho_B$. The normalization coefficient $P_g(m)=\sum_{n=m}^N|\lambda_{n}(\tau)|^2p_{n-1}^{(m-1)}$ is the measurement probability of the $m$th round. In Eq.~(\ref{PwOnChargingrhoB}), a projective measurement generates a population transfer between neighboring energy ladders of the battery $p_{n}^{(m)}\leftarrow|\lambda_{n}(\tau)|^2p_{n-1}^{(m-1)}$ with an $n$-dependent weight $|\lambda_{n}(\tau)|^2$. The battery is initially set as a Gibbs thermal state with populations following an exponential decay function of the occupied-state index $n$. Thus it is charged step by step by the POVM $\mathcal{M}_{eg}$, which moves the populations of lower-energy states up to higher-energy states. And the $\tau$-dependent normalization coefficient $|\lambda_{n}(\tau)|^2$ ranges from zero to unit. $p_{n}^{(m)}$ is thus determined by the measurement interval $\tau$ between two consecutive measurements. With a sequence of properly designed or optimized measurement intervals, the battery could be constantly charged by the projection-induced POVM.

\begin{figure}[htbp]
\centering
\includegraphics[width=0.95\linewidth]{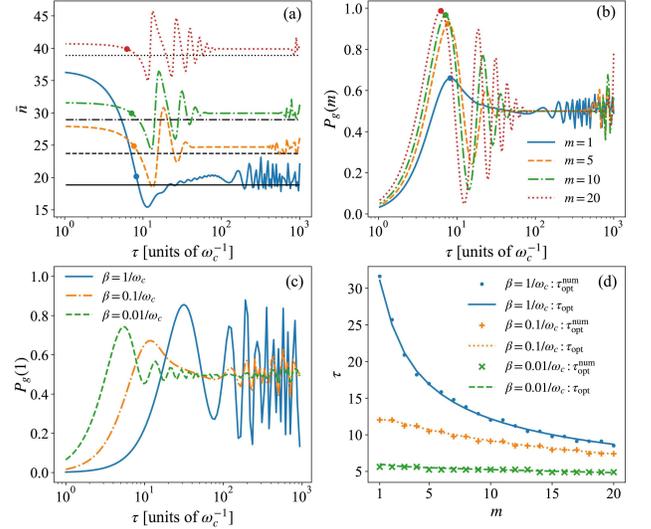}
\caption{(a) Average population $\bar{n}$ and (b) Measurement probability $P_g$ as functions of measurement interval $\tau$ after $m$ rounds of measurements within the power-on scheme. The black lines in (a) represent average populations $\bar{n}^{(m-1)}$ after $(m-1)$ measurements, among which the black solid line denotes the initial thermal occupation $\bar{n}_{\rm th}$ of the battery. The closed circles in (a) represent the moments for the $m$th measurement as obtained from (b) with a maximal measurement probability, by which the updated $\bar{n}^{(m)}$ is found to be larger than $\bar{n}^{(m-1)}$, promising a charging with a significant probability. (c) Measurement probability as a function of $\tau$ under various temperatures. (d) Sequences of the optimized measurement intervals under various temperatures, where $\tau_{\rm opt}$ is given by Eq.~(\ref{OptTau}) and $\tau_{\rm opt}^{\rm num}$ represents the numerical result. The battery size is $N=100$, the initial inverse temperature in (a) and (b) is $\beta=0.05/\omega_c$, the detuning between charger qubit and battery is $\Delta/\omega_c=0.02$, and the coupling strength is $g/\omega_c=0.04$. }\label{Interval_Pon}
\end{figure}

We present the average population $\bar{n}^{(m)}\equiv\sum_nnp_n^{(m)}$ after $m$ measurements in Fig.~\ref{Interval_Pon}(a) as a function of measurement interval $\tau$. The initial thermal population $\bar{n}_{\rm th}$ is plotted (see the black solid line) to compare with $\bar{n}^{(1)}$. One can find a considerable charging effect in the range of measurement interval $\tau\leq9/\omega_c$. Over this critical point, a cooling-effect range appears where the average population is less than the initial population. And afterwards $\bar{n}$ fluctuates with an even larger $\tau$. To pursue the highest charging power, one might intuitively choose a measurement interval as small as possible. It is however under the constraint of a practical coupling strength between charger qubits and battery. In addition, if the measurement interval is smaller than the characterized period of the charger qubit ($\tau\sim1/\omega_c$ or less), then the joint-evolution interval would be too short to detect the charger qubit (initially in the excited state) in its ground state. It will extremely suppress the measurement probability.

As the measurement probability $P_g(1)$ shown by the blue solid line in Fig.~\ref{Interval_Pon}(b), it approaches zero when $\tau\omega_c\rightarrow0$. When the measurement interval approaches about $\tau=8/\omega_c$, $P_g(1)$ climbs to a peak value over $68\%$ and then declines with $\tau$ and ends up with a random fluctuation. It is interesting and important to find that there is a mismatch between the charging-discharging critical point of $\bar{n}$ (the crossing between the black solid line and the blue solid line) in Fig.~\ref{Interval_Pon}(a) and the peak value of $P_g(1)$. With the optimized measurement interval for the maximized $P_g(1)$, a single measurement on the charger qubit could enhance the battery energy from about $19\omega_b$ to $20\omega_b$. In the mean time, the charger qubit that is prepared as the excited state and measured on the ground state has achieved the maximum efficiency with respect to the energy transfer during each charging round. Then the rest lines in Figs.~\ref{Interval_Pon}(a) and \ref{Interval_Pon}(b) support that the battery would be constantly charged with a significant probability when the measurement intervals for the ensued rounds of evolution-and-measurement can be optimized by maximizing the measurement probability. For various numbers of measurements, each POVM $\mathcal{M}_{eg}(\tau)$ with the maximal measurement probability is found to charge rather than discharge the battery by enhancing $\bar{n}$.

The measurement probability can be approximately expressed by a finite summation involving sine functions $P_g(m)=\sum_{n=m}^N\sin^2(\Omega_{n}\tau)p_{n-1}^{(m-1)}$ with $\Omega_n\approx g\sqrt{n}$ under the resonant or near-resonant condition. It is estimated that for a sufficiently large $N$ and a sufficiently short $\tau$,
\begin{equation}
\begin{aligned}
P_g(m)&=\sum_{n=m}^N\left[1-\cos^2(\Omega_{n}\tau)\right]p_{n-1}^{(m-1)} \\
&\approx\sum_np_{n-1}^{(m-1)}-\sum_n\cos^2(\Omega_{n}\tau)p_{n-1}^{(m-1)}\\
&=1-\left[1-\left({\Omega_{\bar{n}+1}^{(m-1)}}\tau\right)^2+\cdots\right]\\
&\approx1-\cos^2(\Omega_{\bar{n}+1}\tau),
\end{aligned}
\end{equation}
where $\Omega_{\bar{n}+1}^{(m-1)}=g\sqrt{\bar{n}^{(m-1)}+1}$ is the average Rabi frequency under the resonant condition and $\bar{n}^{(m-1)}\equiv\sum_n(n-1)p_{n-1}^{(m-1)}$ is the average population for the battery after $(m-1)$ rounds of measurements. The optimized measurement interval is thus given by an iterative formula:
\begin{equation}\label{OptTau}
\tau_{\rm opt}^{(m)}=\frac{\pi}{2\Omega_{\bar{n}+1}^{(m-1)}}.
\end{equation}
It means that $\tau_{\rm opt}$ is updated by the battery's average population of the last round. Note the leading-order correction from a nonvanishing detuning $\Delta$ is in its second order.

Equation~(\ref{OptTau}) can be further verified under various temperatures and during multiple rounds of charging. The optimized measurement interval is inversely proportional to the square root of the average population that is roughly inversely proportional to $\beta$. In Fig.~\ref{Interval_Pon}(c), one can find that the overall behaviors of $P_g(1)$ with various $\beta$ are similar to that in Fig.~\ref{Interval_Pon}(b). A bigger $\beta$ yields a larger $\tau_{\rm opt}$ to have a peak value of $P_g(1)$. In other words, one has to perform more frequent measurements to charge a battery initially in a higher temperature. It is also reflected in Fig.~\ref{Interval_Pon}(d), by which we compare the analytical results through Eq.~(\ref{OptTau}) and the numerical results of optimized measurement intervals for $20$ rounds of measurements under various temperatures. It is found that for $\beta$ across three orders in magnitude, the analytical formula~(\ref{OptTau}) is well suited to obtain the maximized measurement probability that represents the maximum energy input from the charger qubits. The effective temperature of battery increases during the charging process. $\tau_{\rm opt}$ then gradually decreases with $m$. It is consistent with the fact that coupling a charger qubit to a higher temperature battery with uniform energy spacing between ladders induces a faster transition between the excited state and the ground state of the qubit.

We have two remarks about the charging efficiency in our measurement-based scheme. First, a decreasing $\tau_{\rm opt}$ with $m$ could give rise to an increasing charging power
\begin{equation}\label{ChargingPowerOn}
\mathcal{P}(m)\equiv\frac{{\rm Tr}\left[H_B\left(\rho_{\rm on}^{(m)}-\rho_{\rm on}^{(m-1)}\right)\right]}{\tau_{\rm opt}^{(m)}},
\end{equation}
which describes the amount of energy accumulated per unit time in a charging round. It is found that such a battery would be charged faster and faster during the first stage of charging process. Second, Fig.~\ref{Interval_Pon}(d) indicates that the time-varying optimized intervals experience dramatic changes in the first several rounds and then become almost invariant as the measurements are implemented. It holds back the time-scales between neighboring projection operations from being too small to lose the experimental feasibility.

According to Eq.~(\ref{PwOnChargingrhoB}), the average population of the battery after $m$ measurements under the resonant or near-resonant condition is
\begin{equation}\label{nbar}
\bar{n}^{(m)}=\frac{\sum_{n=m}^Nn\sin^2(\Omega_n\tau)p_{n-1}^{(m-1)}}{\sum_{n=m}^N\sin^2(\Omega_n\tau)p_{n-1}^{(m-1)}}.
\end{equation}
Around $n'=\bar{n}^{(m-1)}+1$, we have
\begin{equation}
\sin^2(\Omega_n\tau)=\sin^2(\Omega_{n'}\tau)+\frac{g\tau\sin^2(2\Omega_{n'}\tau)}{2\sqrt{n'}}(n-n')+\cdots.
\end{equation}
All of these squares of sine functions could be approximate to the second order of $(n-n')$ as $\sin^2\Omega_n\tau\approx\sin^2\Omega_{\bar{n}+1}^{(m-1)}\tau_{\rm opt}^{(m)}=1$ when each measurement is implemented with the optimal spacing in Eg.~(\ref{OptTau}). Therefore, Eq.~(\ref{nbar}) could be expressed with the average population of the last charging round $\bar{n}^{(m-1)}$:
\begin{equation}\label{linear}
\begin{aligned}
\bar{n}^{(m)}&\approx\frac{\sum_{n=m}^N(n-1)p_{n-1}^{(m-1)}+\sum_{n=m}^Np_{n-1}^{(m-1)}}{\sum_{n=m}^Np_{n-1}^{(m-1)}}\\
&\approx\bar{n}^{(m-1)}+1.
\end{aligned}
\end{equation}
It is assumed that after a sufficient number of measurements, $\sum_{n=m}^Np_{n-1}^{(m-1)}\approx1$. Due to Eq.~(\ref{linear}), the battery takes an almost $100\%$ unit of energy from the charger qubit in each round of evolution-and-measurement. In other words, POVM $\mathcal{M}_{eg}$ promotes a near-perfect charging protocol, whose efficiency overwhelms the charging schemes without measurements. In the charging scheme based on the quantum collision framework~\cite{CollisionalCharging}, the battery takes about $25\%\sim50\%$ unit of energy from the charger qubit in each cycle. Then more charging cycles are demanded to charge the same amount of energy to battery.

\begin{figure}[htbp]
\centering
\includegraphics[width=0.95\linewidth]{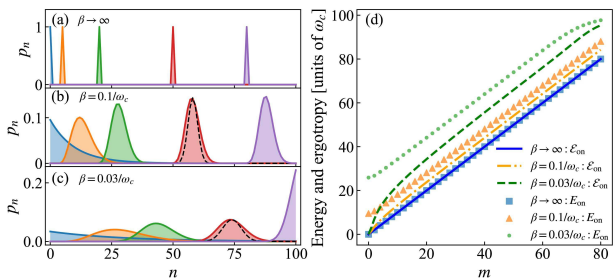}
\caption{Histogram of the battery populations after $m=0$ (blue), $m=5$ (orange), $m=20$ (green), $m=50$ (red), and $m=80$ (purple) measurements for various initial temperatures: (a) $\beta\rightarrow\infty$, (b) $\beta=0.1/\omega_c$, and (c) $\beta=0.03/\omega_c$ under the power-on charging scheme. The black-dashed curves in (b) and (c) describe the near-Gaussian distributions with the same average population and variance of the battery state after $m=50$ measurements under finite temperatures. (d) Battery ergotropy (lines) and energy (makers) as functions of $m$ under various temperatures. The other parameters are the same as those in Fig.~\ref{Interval_Pon}. }\label{PwOnChargingFig}
\end{figure}

Consider the ground-state or zero-temperature case $\rho_B=|0\rangle\langle0|$ as described by the blue histogram in Fig.~\ref{PwOnChargingFig}(a), which is the ``easy mode'' in previous schemes of quantum battery. Each POVM $\mathcal{M}_{eg}$ allows a full state transfer from lower to higher levels $p_{n+1}^{(m)}\leftarrow p_{n}^{(m-1)}$. After $m<N$ measurements, the whole population is transferred to the $m$th energy ladder of the battery with zero variance
\begin{equation}
\Delta n^{(m)}\equiv\sum_nn^2p_n^{(m)}-\left[\bar{n}^{(m)}\right]^2=0.
\end{equation}
And in this case, the charged energy
\begin{equation}\label{batteryenergy}
E_{\rm on}^{(m)}\equiv{\rm Tr}[H_B\rho_{\rm on}^{(m)}]
\end{equation}	
and the ergotropy
\begin{equation}\label{batteryergotropy}
\mathcal{E}_{\rm on}^{(m)}=E_{\rm on}^{(m)}-{\rm Tr}[\sigma_{\rm on}^{(m)}H_B]
\end{equation}	
are exactly the same. Here $\sigma_{\rm on}^{(m)}$ is the passive state of $\rho_{\rm on}^{(m)}$. Thus the power-on charging scheme realizes a full conversion between excitations of charger qubits and usable energy of the battery, when the latter starts from a pure Fock state. This result can be intuitively obtained by a scheme based on the energy swap operations between charger qubit and battery. However, it is hardly extended to more practical scenario for arbitrary states of both charger qubit and battery.

In Fig.~\ref{PwOnChargingFig}(b), the battery is prepared with a moderate temperature. The battery state is gradually transformed from a thermal distribution to a Gaussian-like one with increasing mean value under measurements. For the battery state $\rho_{\rm on}^{(m)}$ with $m=50$ measurements (see the red histogram), a Gaussian state $\rho_G$ with the same average population and variance is distinguished with a black dashed curve. The fidelity $F(m)={\rm Tr}\sqrt{\sqrt{\rho_G}\rho_{\rm on}^{(m)}\sqrt{\rho_G}}$ between them is found to be $0.82$ when $m=50$. Analogous to the Fano factor~\cite{FanoFactor}, we can also use the ratio of the variance and the mean value $f(m)\equiv \Delta n^{(m)}/\bar{n}^{(m)}$ to characterize the evolution of the population histograms. It is found that $f(5)=1.37$, $f(20)=0.33$, $f(50)=0.13$, and $f(80)=0.08$. As measurements are constantly implemented, the battery state distribution thus becomes even sharper and the populations tend to concentrate around $\bar{n}$, providing more extractable energy.

The varying histograms under a higher temperature are plotted in Fig.~\ref{PwOnChargingFig}(c), where the fidelity $F(m=50)$ is still about $0.82$ and the variance is much extended in Fock space. Comparing the purple distributions in Figs.~\ref{PwOnChargingFig}(c) and \ref{PwOnChargingFig}(b), it is more easier for a higher-temperature battery gives rise to a population inversion than a lower-temperature one, as the measurement number approaches the battery size $m\sim N$.

We demonstrate the energy and ergotropy of the battery as functions of the measurement number in Fig.~\ref{PwOnChargingFig}(d). It is interesting to find that the finite temperature does not constitute an obstacle of the power-on scheme in achieving a near-unit ratio of ergotropy and energy, as presented in the case of zero temperature. Both energy and ergotropy scale linearly as indicated by Eq.~(\ref{linear}) with the number of POVMs. It means that our power-on scheme is capable to realize a near-unit rate of energy transfer and achieve a high-ergotropy state, without preparing the battery as a Fock state~\cite{CollisionalCharging}. No longer the thermal state is a ``hard mode'' for quantum battery. Before the occurrence of population inversion, the relative amount of the unusable energy for a lower-temperature battery is larger than a higher-temperature one, e.g., when $m=60$, we have $\mathcal{E}_{\rm on}^{(m)}/E_{\rm on}^{(m)}\approx0.94$ for $\beta=0.1/\omega_c$ and $\mathcal{E}_{\rm on}^{(m)}/E_{\rm on}^{(m)}\approx0.91$ for $\beta=0.03/\omega_c$. In contrast, when $m=80$, we have $\mathcal{E}_{\rm on}^{(m)}/E_{\rm on}^{(m)}\approx0.96$ and $0.98$, for $\beta=0.1/\omega_c$ and $0.03/\omega_c$, respectively. It is reasonable since population inversion indicates a close-to-unit utilization ratio of ergotropy and energy. In the collision model without measurement~\cite{CollisionalCharging}, multiple times of $N$ rounds of cycles are required to achieve the same high ratio.

\begin{figure}[htbp]
\centering
\includegraphics[width=0.95\linewidth]{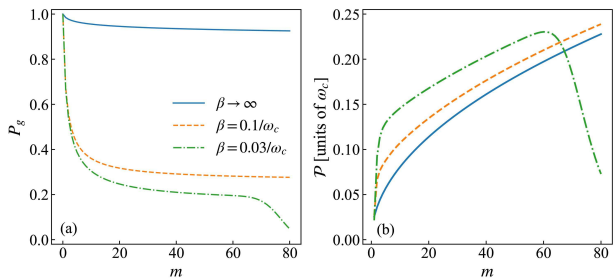}
\caption{(a) Success probability and (b) Charging power under power-on charging as functions of measurement number. All parameters are the same as Fig.~\ref{Interval_Pon} apart from the inverse temperature.}\label{PgAndPower}
\end{figure}

The success probabilities of the power-on charging scheme under various temperatures are plotted in Fig.~\ref{PgAndPower}(a), which is defined as a product of the measurement probabilities of all the rounds $P_g=\prod_mP_g(m)$. For charging the battery at the vacuum state ($\beta\rightarrow\infty$), the success probability decreases with a very small rate. It is still over $93\%$ when $m=80$. When charging a finite-temperature thermal state, the success probability experiences an obvious decay during the first several ($m\approx20$) rounds of measurements. Then the histogram of battery population is transformed to be a near-Gaussian distribution [see Fig.~\ref{PwOnChargingFig}(b) and \ref{PwOnChargingFig}(c)] and it becomes sharper as more measurements are implemented. It gives rise to $P_g(20\leq m\leq60)\approx1$ and thus $P_g$ is almost invariant before the occurrence of population inversion ($m\approx60$). For a moderate temperature $\beta=0.1/\omega_c$, the battery would be successfully charged with a $28\%$ probability under $m=80$ measurements. For a higher temperature $\beta=0.03/\omega_c$, the success probability declines from about $20\%$ to $5\%$ as $m=60\rightarrow m=80$. During the last stage, the upperbound level of the battery is populated and then a larger portion of the near-Gaussian distribution of population has to be abandoned under normalization as more measurements are performed. It is thus expected to have a decreasing $P_g(m>60)$.

The charging power defined in Eq.~(\ref{ChargingPowerOn}) can be observed in Fig.~\ref{PgAndPower}(b), also exhibiting a similar monotonic pattern under various temperatures in a large range. The average population increases linearly with the measurement number and the optimized measurement interval is inversely proportional to the square root of the average population according to Eqs.~(\ref{linear}) and (\ref{OptTau}), respectively. Then it is found that the charging power increases approximately as $\mathcal{P}(m)\varpropto\sqrt{\bar{n}^{(m)}}$. As shown in Fig.~\ref{PgAndPower}(b), a higher temperature gives rise to a larger $\mathcal{P}(m)$ until a decline behavior after $m=60$ measurements (see the green dot-dashed line). That behavior is also induced by the population inversion, on which the average population of the battery fails to keep a linear growth as indicated by the green dotted line in Fig.~\ref{PwOnChargingFig}(d).

\section{Power-off charging}\label{PwoffChargingSec}

When the measurement basis does not commute with the system Hamiltonian, it allows to take the energy away from the measurement apparatus and deposit it to the system. In such a way, energy turns to be useful work~\cite{MeasurementEngines}. When the charger qubits are not in their excited state, getting the state information from them by projection-induced POVMs can convert the information to usable energy through work done on the battery~\cite{Jacobs2009}. And the energy cost of a measurement depends on the work value of the acquired information~\cite{Jacobs2012}.

We now analyse the power-off charging scheme described by $\mathcal{M}_{ge}$, in which charger qubits are prepared in their ground state with $q=1$, and projective measurement is performed on the excited state with $\theta=\pi$ (see the upper right corner in Fig.~\ref{ChargingDischarging}). In this case, the energy change in the charger-battery system is caused by the measurements alone. The battery state after $m$ rounds of power-off charging reads
\begin{equation}\label{PwOffChargingrhoB}
\rho_{\rm off}^{(m)}=\frac{\mathcal{M}_{ge}\left[\rho_{\rm off}^{(m-1)}\right]}{P_e(m)}=\frac{\sum_{n=0}^{N-1}|\lambda_{n+1}(\tau)|^2p_{n+1}^{(m-1)}|n\rangle \langle n|}{P_e(m)}
\end{equation}
with $P_e(m)=\sum_{n=0}^{N-1}|\lambda_{n+1}|^2p_{n+1}^{(m-1)}$ the measurement probability of the $m$th round. In contrast to $\mathcal{M}_{eg}$ in Eq.~(\ref{PwOnChargingrhoB}), Eq.~(\ref{PwOffChargingrhoB}) indicates that $\mathcal{M}_{ge}$ replaces the populations on low-energy states with those on their neighboring high-energy states weighted by a $\tau$-dependent factor $p_n^{(m)}\leftarrow|\lambda_{n+1}(\tau)|^2p_{n+1}^{(m-1)}$. For a battery initially in a Gibbs state, whose population is maximized on the ground state, $\mathcal{M}_{ge}$ could have a certain degree of charging effect on the battery after population renormalization by moving a smaller occupation on higher levels to lower levels. The low-energy states are thus always populated during the histogram evolution from a thermal distribution to a near-Gaussian distribution.

\begin{figure}[htbp]
\centering
\includegraphics[width=0.95\linewidth]{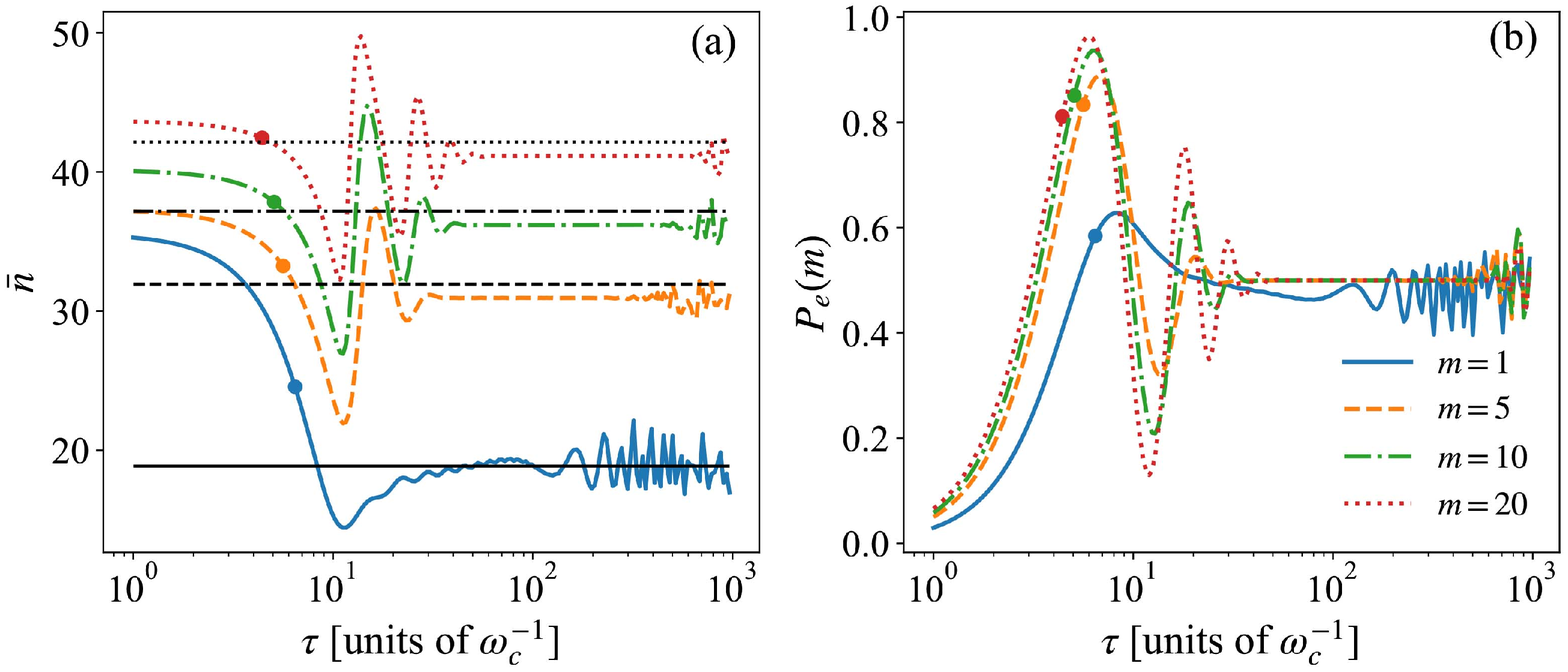}
\caption{(a) Average population $\bar{n}$ and (b) Measurement probability $P_e$ as functions of measurement interval $\tau$ after $m$ rounds of measurements within the power-off scheme. The black lines in (a) represent average populations $\bar{n}^{(m-1)}$ after $(m-1)$ measurements. The closed circles in (a) and (b) represent the compromise result of the changing ratio and the success probability for the moments of the $m$th measurement, by which the updated $\bar{n}^{(m)}$ is maintained over $\bar{n}^{(m-1)}$, promising a charging with a reasonable probability until $m\approx20$. $\beta=0.05/\omega_c$, $\Delta/\omega_c=0.02$, and $g/\omega_c=0.04$.}\label{Interval_Poff}
\end{figure}

The charging effect by the power-off scheme is limited by optimizing the measurement intervals. We provide the average population of battery and the measurement probability under multiple POVMs $\mathcal{M}_{ge}$ as functions of the measurement interval in Figs.~\ref{Interval_Poff}(a) and \ref{Interval_Poff}(b), respectively. Both of them present similar patterns as those under the power-on scheme in Fig.~\ref{Interval_Pon}(a) and \ref{Interval_Pon}(b). However, the mean value of the population $\bar{n}$ is less than its initial value $\bar{n}_{\rm th}$ (indicated by the black solid line) when $\tau$ is chosen such that the measurement probability attains the peak value. In this case, the battery is discharged rather than charged. In general, it is then hard to charge the battery with a significant success probability under a number of rounds of evolution-and-measurement. To charge the battery within the power-off scheme, we have to compromise the charging ratio of neighboring rounds $r=\bar{n}^{(m+1)}/\bar{n}^{(m)}$ and the success probability $P_e=\prod_mP_e(m)$ by numerical optimization. Here we choose to maximize $\exp(xP_e)\log_x{r}$ to ensure the battery could be charged constantly with a reasonable success probability, where $x$ is an index to balance the weights of $r$ and $P_e$ and chosen as $x=10$ in the current simulation. The optimized results for the power-off charging are distinguished by the closed circles in Figs.~\ref{Interval_Poff}(a) and \ref{Interval_Poff}(b). In contrast to Fig.~\ref{Interval_Pon}(b), one has to perform the measurements with a shorter spacing $\tau$ before $P_e(m)$ attains the peak value. The battery could then be constantly charged yet the number of sequential measurements is much limited [see the dotted lines in Fig.~\ref{Interval_Poff}(a) with $m=20$]. The success probability $P_e$ for the power-off scheme is doomed to be much smaller than $P_g$ for the power-on scheme.

\begin{figure}[htbp]
\centering
\includegraphics[width=0.95\linewidth]{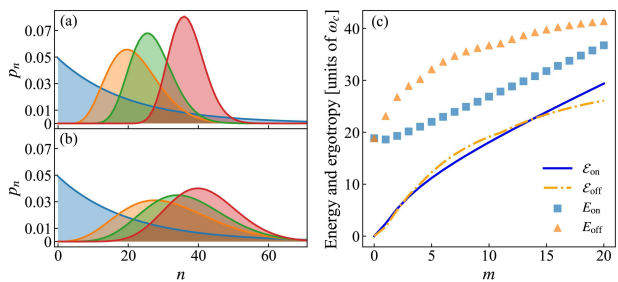}
\caption{Histogram of the battery populations under (a) power-on and (b) power-off charging schemes after $m=0$ (blue), $m=5$ (orange), $m=10$ (green), and $m=20$ (red) measurements. (c) Battery ergotropy (lines) and energy (markers) as function of $m$ under both power-on and power-off schemes. Parameters are the same as those in Fig.~\ref{Interval_Poff}.}\label{PwOffChargingFig}
\end{figure}

To compare the charging efficiency under the power-on and power-off schemes, the evolutions of the battery-population distribution under various number of measurements are demonstrated in Fig.~\ref{PwOffChargingFig}(a) and \ref{PwOffChargingFig}(b), respectively. With the same setting of parameters and initial thermal state, it is found that the the power-on charging and the power-off charging are almost the same in the mean values of battery population, yet are dramatically different in the variances. For example, when $m=20$, $\bar{n}^{(m)}$ are found to be about $37$ and $41$ and $\Delta n^{(m)}$ are about $24$ and $98$ for $\rho_{\rm on}^{(m)}$ and $\rho_{\rm off}^{(m)}$, respectively. The population distribution under the power-off charging is much more extended than the power-on charging, which is not favorable to a quantum battery.

Mean value and variance of the battery population are associated with ergotropy and energy in Fig.~\ref{PwOffChargingFig}(c), where $\mathcal{E}_{\rm off}^{(m)}=E_{\rm off}^{(m)}-{\rm Tr}[\sigma_{\rho_{\rm off}}^{(m)}H_B]$ and $E_{\rm off}^{(m)}={\rm Tr}[H_B\rho_{\rm off}^{(m)}]$ with $\sigma_{\rho_{\rm off}}^{(m)}$ the passive state of $\rho_{\rm off}^{(m)}$. For the power-off charging scheme, the battery cannot obtain energy from both charger qubits and the external energy input apart from projective measurements, as indicated by $p_n$ in Fig.~\ref{PwOffChargingFig}(b) and $E_{\rm off}$ in Fig.~\ref{PwOffChargingFig}(c). The charger qubit is charged in the same time as the battery since the measurement is performed on its excited state. With respect to the battery initial state, the power-off charging scheme can convert the thermal state of zero ergotropy to a high-ergotropy state, especially under less numbers of measurements [see $m=5\to m=10$ in the yellow dot-dashed line of Fig.~\ref{PwOffChargingFig}(c)]. The numerical simulation at $m=20$ shows that the success probabilities for the power-off and power-on schemes are about $1\%$ and $26\%$, respectively. For $m\approx20$, the charged energy under the power-off scheme is larger than that under the power-on scheme. But the latter takes advantage in ergotropy, more crucial for a quantum battery. When $m>15$, the advantage becomes more significant with respect to the ratio of ergotropy and energy.

\section{Discussion}\label{Discussion}

\subsection{Charging in the presence of initial coherence of charger qubits}

\begin{figure}[htbp]
\centering
\includegraphics[width=0.95\linewidth]{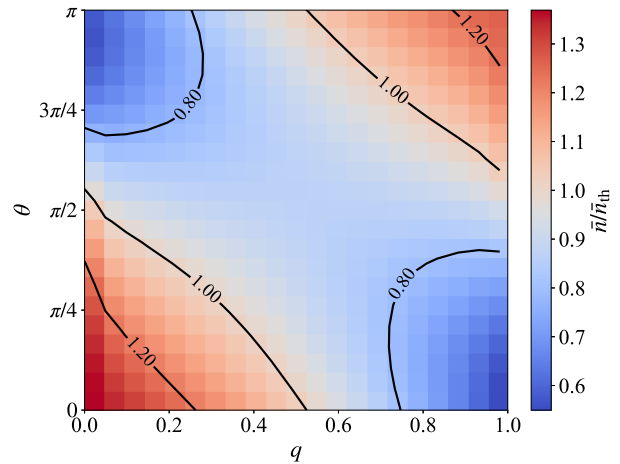}
\caption{Ratio $\bar{n}/\bar{n}_{\rm th}$ of the average population $\bar{n}$ after a single measurement on the battery and the initial thermal population $\bar{n}_{\rm th}$ in the space of $\theta$ and $q$ with full initial coherence $c=1$. Parameters are set as the same as those in Fig.~\ref{ChargingDischarging}.}\label{ChargingDischarging_coh}
\end{figure}

Rather than the dynamical coherence in Eq.~(\ref{DynCoherence}) that is suppressed by the projective measurements, we can discuss the effect from the initial coherence, when the charger qubit is initialled as $\rho_C=q|g\rangle\langle g|+(1-q)|e\rangle\langle e|+c\sqrt{q(1-q)}(|e\rangle\langle g|+|g\rangle\langle e|)$. Here $0\leq c\leq1$ serves as a coherence indicator. When $c=0$, it recovers the preceding analysis in Sec.~\ref{MeasureFueledCharger}. When $c=1$, the charger qubit is in a superposed state. Consequently, it is found that the population of the battery state in Eq.~(\ref{diag}) becomes
\begin{equation}
\mathcal{D}=\mathcal{D}_{\rm charge}+\mathcal{D}_{\rm discharge}+\mathcal{D}_c,
\end{equation}
where $\mathcal{D}_{\rm charge}$ and $\mathcal{D}_{\rm discharge}$ are the same as Eq.~(\ref{Population}) and
\begin{equation}
\begin{aligned}
\mathcal{D}_c=&c\sqrt{q(1-q)}\sin\theta\sum_{n=0}\Big[\cos(\Omega_n\tau)\cos(\Omega_{n+1}\tau)\\
&-\frac{\Delta^2}{4\Omega_n\Omega_{n+1}}\sin(\Omega_n\tau)\sin(\Omega_{n+1}\tau)\Big]p_n|n\rangle\langle n|
\end{aligned}
\end{equation}
represents the population contributed from coherence. Under the approximation $\Omega_{n+1}\approx\Omega_n$ and the near-resonant condition $\Delta\approx0$, we have
\begin{equation}\label{DcApprox}
\mathcal{D}_c\approx c\sqrt{q(1-q)}\sin\theta\mathcal{M}_{gg}[\rho_B],
\end{equation}
where $\mathcal{M}_{gg}$ is the POVM defined in Eq.~(\ref{POVM}). As indicated by Fig.~\ref{ChargingDischarging}, the initial coherence then prefers to cool down rather than to charge the battery.

The negative role played by the initial coherence in charging can be confirmed by Fig.~\ref{ChargingDischarging_coh} about the average population ratio $\bar{n}/\bar{n}_{\rm th}$ as a function of the measurement parameter $\theta$ and the initial state parameter $q$ when $c=1$. In comparison to Fig.~\ref{ChargingDischarging} without the initial coherence, it is found that the effects from different POVMs remain invariant. $\mathcal{M}_{eg}$ (the lower left corner) and $\mathcal{M}_{ge}$ (the upper right corner) still dominate the charging effects and $\mathcal{M}_{ee}$ (the upper left corner) and $\mathcal{M}_{gg}$ (the lower right corner) are still in charge of discharging. While the former becomes weakened with less red area and the latter becomes enhanced with more blue area. The numerical result is consistent with Eq.~(\ref{DcApprox}).

\subsection{Charging in the presence of decoherence}

\begin{figure}[htbp]
\centering
\includegraphics[width=0.95\linewidth]{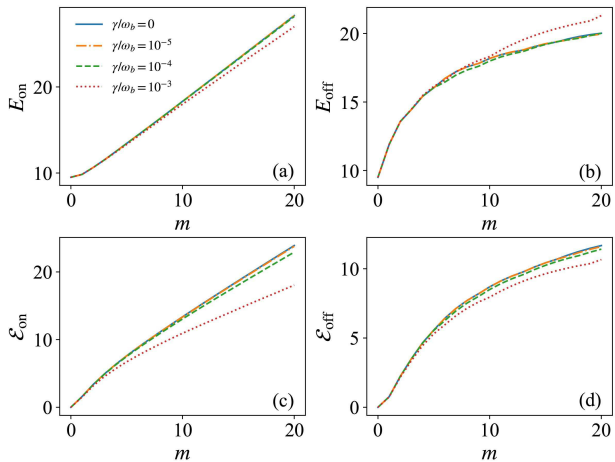}
\caption{(a) and (b): Battery energy as a function of measurement number $m$ with various dissipation rates $\gamma$ for the power-on and power-off charging schemes, respectively. (c) and (d): Battery ergotropy as a function of $m$ with various dissipation rates under the power-on and power-off charging schemes, respectively. The initial inverse temperature is $\beta=0.1/\omega_c$. The other parameters are the same as Fig.~\ref{Interval_Pon}.}\label{OpenSystem}
\end{figure}

In realistic situations, any charging process should be considered in an open-quantum-system scenario. We now estimate the impact from the environmental decoherence on our charging schemes based on the measurements. The dynamics of the full system can then be described by the master equation,
\begin{equation}\label{ME}
\begin{aligned}
&\dot{\rho}_{\rm tot}(t)=-i[H, \rho_{\rm tot}(t)]\\
&+\gamma_b(\bar{n}_{\rm th}+1)\mathcal{D}[A]\rho_{\rm tot}(t)+\gamma_b\bar{n}_{\rm th}\mathcal{D}[A^\dagger]\rho_{\rm tot}(t)\\ &+\gamma_c(\bar{n}_{\rm th}^c+1)\mathcal{D}[\sigma_-]\rho_{\rm tot}(t)+\gamma_c\bar{n}_{\rm th}^c\mathcal{D}[\sigma_+]\rho_{\rm tot}(t),
\end{aligned}
\end{equation}
where $H$ is given by Eq.~(\ref{Ham}) and $\mathcal{D}[o]$ represents the Lindblad superoperator $\mathcal{D}[o]\rho_{\rm tot}(t)\equiv o\rho_{\rm tot}(t)o^\dagger-1/2\{o^\dagger o, \rho_{\rm tot}(t)\}$. Here $\gamma_b$ and $\gamma_c$ are dissipative rates for battery and charger qubits, respectively. $\bar{n}_{\rm th}={\rm Tr}[\hat{n}\rho_B]$ and $\bar{n}_{\rm th}^c=1-q_{\rm th}$ are their initial thermal average occupations, where $q_{\rm th}=1/(e^{\beta\omega_b}+1)$ is the thermal population on the ground state of charger qubits.

Figure~\ref{OpenSystem} presents both energy and ergotropy of the battery under two charging schemes with various dissipative rates along a small sequence of $m\leq20$ charging rounds. It is demonstrated that in the presence of the thermal decoherence with $\gamma_b=\gamma_c=\gamma\leq10^{-4}\omega_b$, both energy and ergotropy deviate slightly from the dissipation-free situation [see the blue lines or Fig.~\ref{PwOffChargingFig}(b)]. It is reasonable because the optimized measurement interval $\tau_{\rm opt}$ decreases as more measurements implemented [see Fig.~\ref{Interval_Pon}(d)], which significantly reduces the environmental effect.

\section{Conclusion}\label{Conclusion}

We established for quantum battery a charging-by-measurement framework based on rounds of joint-evolution and partial-projection. The charger system is constituted by a sequence of disposable qubits. General POVMs on the battery system of $N+1$ levels are constructed by the exchange interaction between charger and battery and the projective measurement on charger qubits in a general mixed state. In the absence of initial coherence, we focus on the charging effect by POVM alone. Despite the battery starts from the thermal-equilibrium state as a ``hard mode'' for quantum battery, it is found that a considerable charging effect can be induced when the qubit is prepared at the excited state and measured on the ground state or in the opposite situation. They are termed as power-on and power-off charging schemes. The power-on scheme exhibits great advantages in the charging efficiency over the schemes without measurements. Under less than $N$ measurements with optimized intervals, our measurement-based charging could transform the battery from a finite-temperature state to a population-inverted state, holding a near-unit ergotropy-energy ratio and a significant success probability. Within a much less number of measurements than the power-on scheme, the power-off charging scheme can be used to charge the battery and charger qubits without external energy input, although it is hard to survive more rounds of measurements with a finite probability.

The POVM in our work manifests a powerful control tool to reshape the population distribution of the battery system, building up a close relation with the ergotropy. Our work therefore demonstrates that quantum measurement can become a useful thermodynamical resource analogous to conventional heat or work reservoirs, serving as high-efficient fuels powering a charging scheme through a nonunitary procedure.

\section*{Acknowledgments}

We acknowledge financial support from the National Science Foundation of China (Grants No. 11974311 and No. U1801661).

\bibliographystyle{apsrevlong}
\bibliography{ref}

\end{document}